\documentclass{epl}

\usepackage{graphicx}
\usepackage{dcolumn}
\usepackage{bm}
\usepackage{bbm}
\usepackage{psfrag}
\usepackage{subfigure}
\usepackage{multirow,amssymb,amsbsy,amsmath}
\usepackage{stmaryrd}
\usepackage{subfigure}

\def\>{\rangle}
\def\<{\langle}
\def\n{\nonumber}

\def\<{\langle}
\def\>{\rangle}
\def\be#1\ee{\begin{equation}#1\end{equation}}
\def\ba{\begin{eqnarray}}
\def\ea{\end{eqnarray}}
\newcommand*{\Haver}[1]{\mathopen{\llbracket} #1 \mathclose{\rrbracket}}

\title{Thermalization of quantum systems by finite baths}
\shorttitle{Thermalization of quantum systems}
\author{Jochen Gemmer\inst{1}\thanks{\email{jgemmer@uos.de}}%
        \and Mathias Michel\inst{2}}
\shortauthor{J. Gemmer et al.}
\institute{
  \inst{1} Physics Department, University of Osnabr\"uck, %
           Barbarastr.\ 7, 49069 Osnabr\"uck, Germany \\%
  \inst{2} Institute of Theoretical Physics I, University of Stuttgart, %
           Pfaffenwaldring 57, 70550 Stuttgart, Germany%
}

\begin{document}

\maketitle

\begin{abstract}
We consider a discrete quantum system coupled to a finite bath, which may consist of only one particle, in contrast to the standard baths which usually consist of continua of oscillators, spins, etc. 
We find that such finite baths may nevertheless equilibrate the system though not necessarily in the way predicted by standard open system techniques. 
This behavior results regardless of the initial state being correlated or not.
\end{abstract}

%
%

Due to the linearity of the Schr\"odinger equation concepts like ergodicity or mixing are strictly speaking absent in quantum mechanics. 
Hence the tendency towards equilibrium is not easy to explain. 
However, except for some ideas \cite{Neumann1929,Landau1980} the approaches to thermalization in the quantum domain seem to be centered around the idea of a thermostat, i.e., some environmental quantum system (bath, reservoir), enforcing equilibrium upon the considered system. 
Usually it is assumed that this bath's classical analogon contains an infinite number of decoupled degrees of freedom.

Theories addressing such scenarios are  the projection operator techniques (time-con\-vo\-lu\-tion\-less, Nakajima Zwanzig), the Born approximation (BA) \cite{Breuer2002} and the path integral technique (Feynman Vernon \cite{Weiss1999}).
The projection operator techniques are exact if all orders of the system-bath interaction strength are taken into account which is practically unfeasible. 
However, assuming weak interactions and accordingly truncating at leading order in the interaction strength (BA) produces  an exponential relaxation behavior (c.f. \cite{Caldeira1985,Makri1999}) whenever the bath consists of an continuum of oscillators, spins, etc. 
The origin of statistical dynamics is routinely based on this scheme, if it breaks down no exponential thermalization can a priori be expected.

We find that this scheme breaks down (i.e.\ the BA produces wrong results) if the bath features a special spectral structure which cannot arise from an uncoupled multitude of subsystems or modes (see below). 
We refer to this type of bath as finite bath. 
This holds true even and especially in the limit of weak coupling and arbitrarily dense bath spectra.

Nevertheless a statistical relaxation behavior can be induced by finite baths.
It simply is not the behavior predicted by the BA. 
Thus the principles of statistical mechanics in some sense apply below the infinite particle number limit and beyond the BA.
 
This also supports the concept of systems being driven towards equilibrium through increasing correlations with their baths \cite{Lubkin1978,Lubkin1993,Zurek1994,GemmerOtte2001,Scarani2002} rather than the idea of system and bath remaining factorizable, which is often attributed to the BA \cite{Weiss1999,Breuer2002}. 
\begin{figure}
  \centering
  \includegraphics[width=5cm]{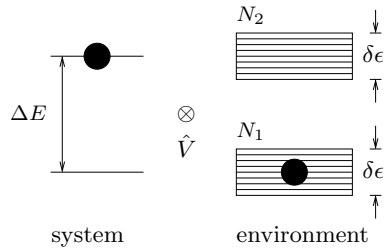}
  \caption{Two-level system coupled to a finite bath. The environment's spectrum deviates significantly from those of baths with infinitely many decoupled degrees of freedom}
  \label{fig:1}
\end{figure}

The model we analyze is characterized (for simplicity) by a two level system (S or ``spin'') coupled two a many level system (B) consisting of two relevant bands featuring the same width and equidistant level spacing (see Fig.~\ref{fig:1}). 
So this may be viewed as a spin coupled to a single molecule, a one particle quantum dot, an atom or simply a single harmonic oscillator. 
Note that the spin, unlike in typical oscillator baths or the Jaynes-Cummings Model, is not in resonance with the environments level spacing but with the energy distance between the bands. 
There are two principal differences of such an  finite environment level scheme from the level scheme of, say, a standard oscillator bath. 
i) The total amount of levels within a band may be finite. 
ii) Even more important, from, e.g., the ground state of a standard bath there are infinitely many resonant transitions to the ``one-excitation-states'' of the bath. 
But from all those, the ``back- transitions'' lead to only one ground state. 
Thus the relevant bands of any infinite bath would consist of only one state in the lower and infinitely many states in the upper band. 
And it will turn out to be that limit in which the standard methods produce correct results. 

A finite bath which cannot be decomposed in uncoupled subunits any further may, however, feature arbitrary numbers of states in both bands. 
Note that in our model there is no notion of the environment being chaotic in itself. 
Due to the considered type of interaction the full system might be termed chaotic, as will become clear below (for a treatment of finite baths under a different perspective, see \cite{Kolovsky1994,Scarani2002}). 
The Hamiltonian of the model in the Schr\"odinger picture reads $\hat{H}=\hat{H}_0+\hat{V}$, $\hat{H}_0$ representing the uncoupled system and $\hat{V}$ the interaction:
\begin{align} 
  \hat{H}_0&=\Delta E \hat{\sigma}_z\n
  +\sum_{n_1}\frac{\delta \epsilon}{N_1}n_1
  |n_1\rangle \langle n_1| +\sum_{n_2}(\Delta E+\frac{\delta \epsilon}{N_2}n_2)
  |n_2\rangle \langle n_2|\n\\
  \hat{V}&=
  \lambda\sum_{n_1,n_2}C(n_1, n_2)\;\hat{\sigma}^{+}|n_1\>\<n_2|
  +\mbox{h.c.}
\end{align}
Here the Pauli matrices refer to S, $n_1(n_2)$ denotes the $n$'th energy eigenstate within the lower(upper) band of B and h.c.\ stands for the hermitian conjugate the previous sum. 
For the example at hand we chose $\Delta E=25u,\; \delta \epsilon=0.5u,\; N_1=N_2=500,\; \lambda=5 \cdot 10^{-4}u$, $u$ being some arbitrary energy unit. 
The real and imaginary parts of the $C$'s are randomly (Gaussian) distributed numbers with mean zero and normalized to $\sum_{n_1,n_2}|C(n_1,n_2)|^2/N_1N_2=1$. 
This interaction type has been chosen in order to keep the model as general and free from peculiarities as possible. 
(In the fields of nuclear physics or quantum chaos random matrices are routinely used to model unknown interaction potentials.
We do, however, analyze the dynamics generated by one single interaction, not the average dynamics of an Gaussian ensemble of interaction matrices.) 

\begin{figure}
\centering
  \subfigure[]{\includegraphics[width=6.5cm]{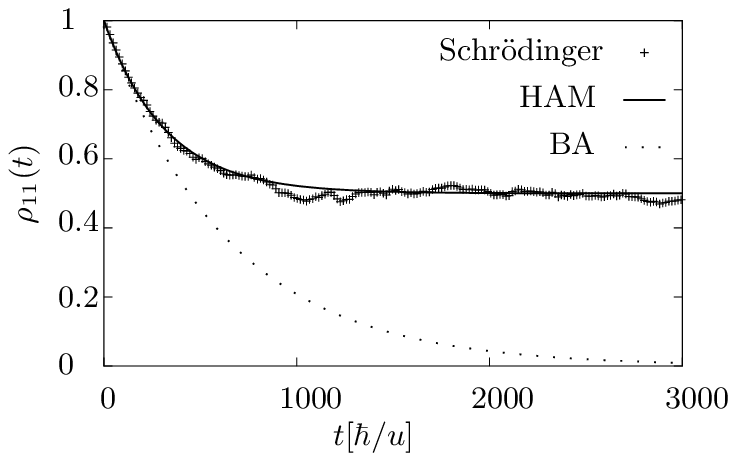}
               \label{fig:2}}\quad
  \subfigure[]{\includegraphics[width=6.5cm]{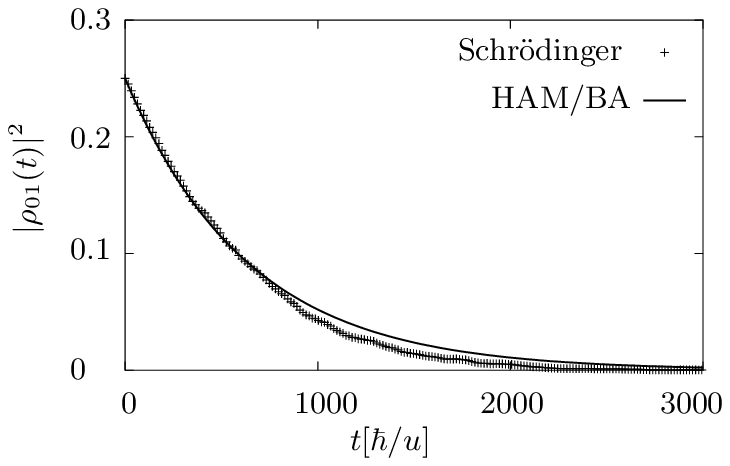}
               \label{fig:3}}
  \caption{Evolution of the excitation probability (product initial state) \textbf{(a)} and the off-diagonal element (correlated initial state) \textbf{(b)} of a spin coupled to a finite bath. Regardless of the baths finiteness the evolution exhibits statistical thermalization respectively coherence vanishes, i.e., the system evolves towards a maximum equilibrium entropy. For the diagonal elements, although, different from what the Born-approximation predicts.}
\end{figure}

We firstly analyze the decay behavior of two different pure product initial states: 
The bath-part of both initial states is a pure state that only occupies the lower band but is, apart from that, chosen at random.
Apart from its pureness only with respect to occupation numbers, B's initial state can be considered an approximation to a Gibbs state with $\delta \epsilon\ll kT_{\text{B}} \ll \Delta E$ (in the example at hand, e.g., $ kT_{\text{B}} \approx 5u$). 
For small $\delta\epsilon$ the temperature may be arbitrarily small. 
Initially, the system S is firstly chosen to be completely in its excited state (this initial state is indicated by the black dots in Fig.~\ref{fig:1}) and, secondly, in a 50:50 superposition of ground and excited state.
The probability (density matrix element $\rho_{11}(t)$) to find the system excited as produced by the first initial state is shown in Fig.~\ref{fig:2}.
Since the first initial state does not contain any off-diagonal elements, we find $|\rho_{01}|^2\approx 0$ for all times.
This is different for the second initial state investigated in Fig.~\ref{fig:3}, it starts with $|\rho_{01}|^2=0.25$ and is thus well suited to study the decay of the coherence.
(The diagonal elements of the second state are already at their equilibrium value ($\rho_{11}(0)=0.5$) in the beginning and exhibit no further change.)

By numerically solving the time-dependent Schr\"odinger equation for the full model's pure state $|\Psi(t)\>$ we find for the reduced state of the system $\hat{\rho}(t)=\mbox{Tr}_{\text{B}} \{|\Psi(t)\>\<\Psi(t)|\}$, an  exponential decay, up to some fluctuations. 
(For the baths initial state being a real mixed Gibbs state one can even expect fluctuations to be smaller, since fluctuations corresponding to various pure addends of the Gibbs state will partially cancel each other.)
Thus the Schr\"odingerian dynamics yield a local behavior that might be described as statistical, controlled by some transition rate.

The full model is Markovian in the sense that bath correlations decay much faster than the system relaxes, concretely bath correlations decay on a time scale of $\tau_c \approx \hbar /\delta \epsilon =2$ (all times given in units of $\hbar/u$), whereas the system relaxes on a timescale $\tau_r \approx 640$ (cf.\ Fig.'s~\ref{fig:2},(b)). 
Since the whole system is finite there is a finite (quasi) recurrence time. 
But due to the incommensurability of the full system's frequencies it appears to be $>10 ^8$, i.e., orders of magnitudes larger than the relaxation time of S.
(The special example at hand features, due to the environments equidistant level spacing, a recurrence time for the bath correlations of approximately $6 \cdot 10^3$ but that does not induce a recurrence in S.)
 
Although the model is Markovian in the above sense and its relaxation appears locally statistical, S's excitation probability deviates significantly from what the BA predicts (cf. Fig \ref{fig:2}): 
The beginning is described correctly, but rather than ending up at
$T=T_\text{B}$ as the BA  predicts for thermal environment states \cite{Breuer2002}, S ends up at $T=\infty$, i.e., equal occupation probabilities for both levels. 
Furthermore a condition often attributed to the BA, namely that S and B remain unentangled, is not fulfilled: 
When S has reached equilibrium the full system is in a superposition of $|$S in the excited state $\otimes$ B in the lower band$\>$ and  $|$S in the ground state $\otimes$ B in the upper band$\>$. 
This is a maximum entangled state with two orthogonal addends, one of which features a bath population corresponding to $T_\text{B}\approx 0$, the other a bath population inversion, i.e., even a negative bath temperature. 
These findings contradict the concept of factorizability but are in accord with a result from \cite{Gemmer2005} claiming that an evolution towards local equilibrium is always accompanied by an increase of system-bath correlations. 
However, the off-diagonal element evolution coincides with the behavior predicted by the BA. 
Thus in spite of the systems finiteness and the reversibility of the underlying Schr\"odinger equation S evolves towards maximum local von Neumann entropy (see Fig.~\ref{fig:3}) which supports the concepts of \cite{Lubkin1993}.

\begin{figure}
\centering
  \subfigure[]{\includegraphics[width=6.5cm]{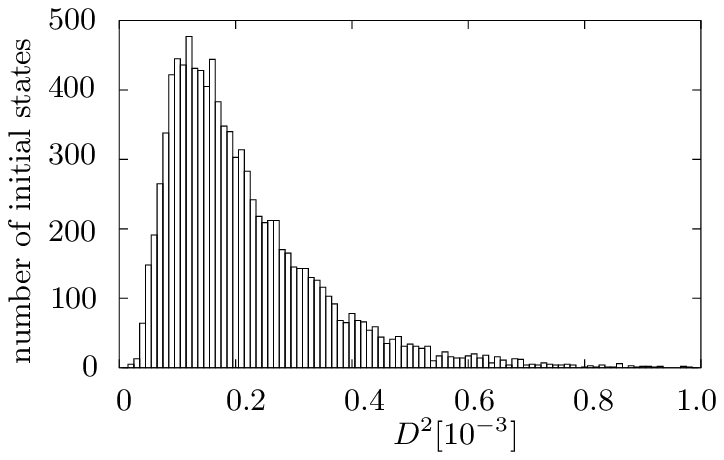}
               \label{fig:4}}
  \subfigure[]{\includegraphics[width=6.5cm]{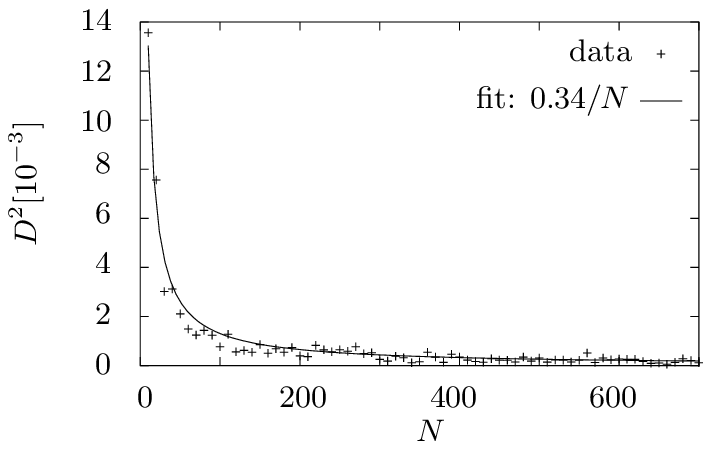}
               \label{fig:5}}
\caption{\textbf{a)} Deviation of the exact evolution of the spins excitation probability from the HAM prediction for a set of entangled initial states. \textbf{b)} Deviation of the exact evolution of the spins excitation probability from the HAM prediction for increasing number $N$ of states in the environment.}
\end{figure}

We now very roughly (and rather incompletely) outline the Hilbert space Average Method (HAM) which explains the behavior of this model-type.
HAM is not limited to two level systems, the example has just been chosen for simplicity. 
(For a detailed description of HAM, see \cite{Gemmer2003,Gemmer2004}). 
We start by considering short time steps of the evolution of S's density matrix $\hat{\rho}$. 
With a truncated (second order) Dyson series for $|\Psi(t+\Delta t)\>\approx \hat{D}(t,\Delta t)|\Psi(t)\>,$ one formally gets
\begin{equation}
  \label{eq:1}
  \hat{\rho}(t+\Delta t)\approx \mbox{Tr}_{\text{B}} \{
  \hat{D}|\Psi(t)\>\<\Psi(t)|\hat{D}^{\dagger} \}.
\end{equation}
If the right hand side of (\ref{eq:1}) (r.h.s.) was only a function of S's local density matrix $\hat{\rho}$, one could set up an iterative scheme for the  local dynamics of S. 
But one finds that the r.h.s.\ depends on all the details of the full model's state $|\Psi(t)\>$ and explicitly on the absolute time $t$, thus no autonomous iterative sub-dynamics for S can be inferred directly. 
However, computing the Hilbert space average ($\Haver{\cdots}$) of the r.h.s., i.e., the average over an adequate set of full system states $|\Phi\>$ sharing (some) crucial quantities with $|\Psi(t)\>$, i.e., $ \<\Phi|\hat{A}|\Phi\>=\<\Psi(t)|\hat{A}|\Psi(t)\>$ with $\hat{A}$ being the full model's energy, the local energy of S or the coherence of S , yields under specific conditions on the model (see below)
\begin{equation}
  \label{eq:2}
  \Haver{\mbox{Tr}_{\text{B}} \{
    \hat{D}|\Phi\>\<\Phi|\hat{D}^{\dagger} \}}  
  \approx \hat{\rho}(t)+\Delta t \left(\mathcal{L}
    \hat{\rho}(t)+\mathcal{I}\hat{\rho}(0)\right), 
\end{equation}
where $\mathcal{L}$, $\mathcal{I}$ are linear super-operators.
(Note that the majority of the states belonging to the above set is correlated, i.e., no factorization is implied.)
Replacing the r.h.s.\ of (\ref{eq:1}) by its Hilbert space average would thus yield an autonomous iteration scheme. 
But is that justified? 
It is justified, whenever (\ref{eq:2}) does not only hold for the Hilbert space average over all $|\Phi\>$'s but also (approximately) for the majority of all $|\Phi\>$'s belonging to the above set individually. 
If this is the case the evolution of S tends to be independent of the full models state and basically controlled by the local state of S. 
This tendency increases with B featuring increasing numbers of eigenstates as can be shown theoretically \cite{Gemmer2004} or numerically (cf.\ Fig.~\ref{fig:5}). 
Nevertheless, this replacement does not represent an approximation with a computable error, it only represents a best unbiased guess for the evolution of S. 
This ``best guess'' structure of HAM accounts for the non-statistical character of the under-laying dynamics, i.e., allows for a non-statistical behavior of S in some (rare) cases. 

Performing the above replacement and taking $\Delta t$ to zero yields for the system-type at hand (but independent of the concrete interaction) the following master equation scheme:
\begin{align}
  \label{eq:6} 
  \dot{\rho}_{11}(t)&=-(R_{10}+R_{01})\rho_{11}(t)+R_{10}\rho_{11}(0)\;, \qquad
  R_{01}=\frac{2\pi\lambda^2N_2}{\hbar \delta \epsilon},\qquad
  R_{10}=\frac{2\pi\lambda^2N_1}{\hbar \delta \epsilon}\n\\
  \dot{\rho}_{01}(t)&=(i\Delta E/\hbar-R_{01}/2)\rho_{01}(t).
\end{align}
The equilibrium value of S's excitation probability is given by $\rho_{11}(\infty)=N_1/(N_1+N_2)$. Thus only if $N_2 \gg N_1$ (infinite bath) the BA produces correct results. 
Otherwise BA must fail for the reduced sub-dynamics of S are not even Markovian now in the sense of the equilibrium state being independent of the initial state. 
Note, however that it is not the finite density of states that causes the break down of the BA, since the BA produces wrong results even for $N_1, N_2 \rightarrow \infty$ as long as the above condition is not met. 
Since boldly calculating transition probabilities according to Fermi's Golden Rule, would have produced the same rates $R$ the applicability of the above scheme implies the applicability of a random phase approximation in the sense of, say, Peierls \cite{Peierls1955}. 

Such an approximation cannot always hold since the Schr\"odinger equation is completely non-random. 
Thus, as necessary conditions on the model parameters for statistically appearing behavior of S we find (from theory and numerics)
\begin{equation}
  \label{eq:7}
  2\lambda\frac{N}{\delta \epsilon}\geq 1, \qquad
  \lambda^2\frac{N}{\delta \epsilon^2}\ll 1,
\end{equation}
where $N$ refers to the band with the larger state density.
Those conditions enforce Markovicity in the above sense, and exclude the infinitely weak coupling limit, $\lambda \rightarrow 0$ as long as B's state density is finite. 
Neither the equidistant level spacing of B nor the Gaussian distribution of the interaction matrix elements are indispensable. 
A condition on the level structure of B is that the number of states within an interval of $I \approx \delta \epsilon /10$ does not depend much on where within the band the interval is chosen. 
A similar condition restricts the interaction: 
The sum of $|C|^2$'s corresponding to transitions from one state of a band to states within an interval $I$ of the other band should not depend much on the position of the interval.

Since HAM is just a ``best guess theory'' the exact evolution follows its predictions with different accuracies for different initial states, even if all conditions on the model are fulfilled.
To analyze this for, say $\rho_{11}(t)$, we introduce $D^2$, being the time-averaged quadratic deviation of HAM from the exact (Schr\"odinger) result 
\begin{equation}  
  \label{eq:12}
  D^2=\frac{1}{\tau}
  \int_0^{\tau}
  \Big(\rho_{11}^{\text{HAM}}(t)-\rho_{11}^{\text{exact}}(t)\Big)^2
  \text{d} t\;.
\end{equation}
Thus $D$ is a measure of the deviations from a predicted behavior.
The results of the investigation for our model (Fig.~\ref{fig:1}) are condensed in the histogram (Fig.~\ref{fig:4}, $\tau=2000$).
The set of respective initial states is characterized by a probability of $3/4$ for $|$S in its excited state $\otimes$ B in its lower band$\>$ and $1/4$ for $|$S in its ground state $\otimes$ B in its upper band$\>$. 
Within these restrictions the initial states are uniformly distributed in the corresponding Hilbert subspace.
Since all of them are correlated the application of a product projection operator technique would practically be unfeasible. 
However, as Fig.~\ref{fig:4} shows, the vast majority of them follows the HAM prediction quite closely, although there is a  typical fluctuation of $D=\sqrt{2}\cdot 10^{-2}$ which is small compared to the features of the predicted behavior (which are on the order of one), due to the finite size of the environment (cf.\ also fluctuations in Fig.~\ref{fig:2}).

In Fig.~\ref{fig:5} the dependence of $D^2$ on the number of states of B is displayed for $N=10,\dots,800$ (one evolution for each environment size).
At $N=500$ like used in the above accuracy investigation we find the same typical fluctuation, whereas for smaller environments the typical deviation is much bigger.
We find that the squared deviation scales as $1/N$ with the environment size, thus making HAM a reasonably reliable guess for many-state environments.
 
What about the claim that reduced dynamics need not to be completely positive, once S and B are correlated \cite{Pechukas1994,Weiss1999,Haenggi2004}? 
In principle this holds. 
But regardless of their being correlated just a small fraction of all states from Fig.~\ref{fig:4}  shows significant deviations from HAM. 
Thus a smooth evolution towards equilibrium (though not necessarily of the Lindblad-type \cite{Alicki1987}) can typically be expected for the reduced dynamics, regardless of the initial state being correlated or not.
\begin{figure}
  \centering
  \includegraphics[width=7cm]{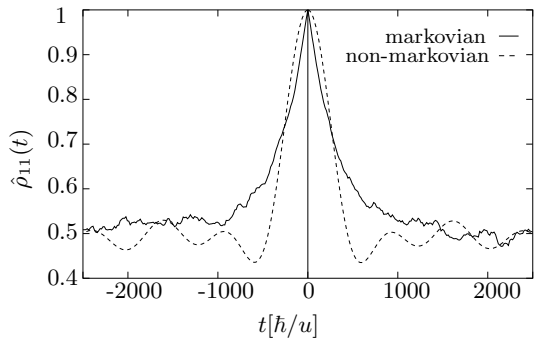}
  \caption{Evolution of the spins excitation probability before and after full excitation for two different sets of model parameters.}
  \label{fig:6}     
\end{figure} 

What about reversibility and statistically appearing dynamics? The Schr\"odinger equation is time reversible and indeed there is no apparent time asymmetry in Fig.~\ref{fig:6}. 
For $t>0$ the behavior of the above system (solid line) is well described by the above rate equation scheme featuring an attractive fix-point. 
But for $t<0$ it would have to be described by a scheme with an repulsive fix-point! 
Thus, any of the full model states at $t<0$ is a paradigm for an initial state that does not yield statistical decay behavior eventhough the model typically generates it. 
The dashed line shows the behavior of a model like the above one, only with $\delta \epsilon \approx 0$ and (in order to keep the timescales comparable) $\lambda =10^{-4}u$. 
This clearly violates the second criterion of (\ref{eq:7}) and the model is no longer Markovian in the above mentioned sense. And indeed even for $t>0$ there is no exponential decay although the model features equally many states as the above one. 
Nevertheless the same equilibrium state as in the above case is reached (cf.\ also \cite{Breuer2004}) regardless of whether the system is propagated forwards or backwards in time.

In essence we have shown that statistical relaxation may emerge directly from the Schr\"o\-dinger equation. 
This requires the respective system being coupled in an adequate way to a suitable environment. 
This environment must feature many eigenstates. 
There is, however, no minimum particle number limit. 
Thus the thermodynamic limit appears to be essentially controlled by the number of environmental eigenstates involved in the dynamics rather than by the number of environmental particles. 
This relaxation behavior results even for correlated initial states, nevertheless, standard open system methods may fail to produce the correct result.

\acknowledgments
We are indebted to  H.-P. Breuer, G. Mahler, A. Kolovsky and A. Buchleitner for interesting discussions on this subject. Financial Support by the Deutsche Forschungsgemeinschaft is gratefully acknowledged.


\end{document}